\documentclass{aa}
\usepackage{psfig}
\usepackage{epsfig}

\def \sax {BeppoSAX}

\def \deg{^\circ}

\def \msun{{\rm ~M_{\odot}}}

\def \hcm {\hbox {\ifmmode $ cm$^{-2}\else cm$^{-2}$\fi}}

\def\approxgt{\mathrel{\hbox{\rlap{\lower.55ex \hbox {$\sim$}}
        \kern-.3em \raise.4ex \hbox{$>$}}}}
\def\approxlt{\mathrel{\hbox{\rlap{\lower.55ex \hbox {$\sim$}}
        \kern-.3em \raise.4ex \hbox{$<$}}}}

\begin{document}

\title{X--ray emission from the giant molecular clouds in
the Galactic Center region and the discovery of new X--ray sources}

\author{L. Sidoli\inst{1}
        \and S. Mereghetti\inst{2}
        \and A. Treves\inst{3}
        \and A.N. Parmar\inst{1}
        \and R. Turolla\inst{4}
	\and F. Favata\inst{1}
}

\offprints{L. Sidoli (lsidoli@astro.estec.esa.nl)}

\institute{
       Astrophysics Division, Space Science Department of ESA, ESTEC,
       Postbus 299, NL-2200 AG Noordwijk, The Netherlands
\and
       Istituto di Fisica Cosmica G.Occhialini,  C.N.R., Via Bassini 15, I-20133 Milano,
       Italy
\and
    Universit\`a degli Studi dell'Insubria, Polo di Como,
    Dipartimento di Scienze Chimiche, Fisiche, e Matematiche, Via Lucini 3, I-22100, Como, Italy
\and
        Universit\`a di Padova, Dipartimento di Fisica, Via Marzolo 8, I-35131, Padova, Italy
}
\date{Received 19 January 2001; Accepted 5 April 2001 }

\authorrunning{L. Sidoli et al.}

\titlerunning{X--rays from Sgr~B, Sgr~ C and Sgr~D} 

\abstract{We report the results of X--ray (2--10~keV) observations of
the giant molecular clouds
Sgr~B, Sgr~C and Sgr~D in the Galactic Center region, 
together with the discovery of the point-like
source SAX~J1748.2--2808.
The data have been obtained with the MECS instrument on the
BeppoSAX satellite.
The core of Sgr~B2 has an X--ray luminosity
of $\sim$6$\times10^{34}$~erg~s$^{-1}$ and its spectrum
is characterized by a strong Fe emission line at $\sim$6.5 keV 
with an
equivalent width of 2~keV.
Faint diffuse X--ray emission is detected from Sgr~C and from the
SNR  G1.05--0.15 (Sgr~D).
A new, unresolved source with a strong Fe line has been
discovered in the Sgr~D region. This source, SAX~J1748.2--2808,
is probably associated with a SiO and OH maser source at the
Galactic Center distance.
If so, its luminosity is 10$^{34}$~erg~s$^{-1}$.
We propose that the 
X--ray emission from SAX J1748.2--2808 
is produced either by protostars or by a giant molecular cloud
core. 
Emission from sources similar to SAX J1748.2--2808  
could  have  an impact
on the expected contribution on
the observed Fe line emission from the Galactic ridge.
} \maketitle

\section{Introduction}

The Galactic Center (GC) region is
characterized by a strong concentration
of point-like  X--ray sources
and  by  intense diffuse
X--ray emission, discovered with the
{\it Einstein} Observatory in the 0.5--4 keV
energy range (Watson et al. 1981).
The $Ginga$ satellite revealed the presence of a 6.7~keV iron line
emission from the Galactic
plane, particularly bright towards the GC direction
(Koyama et al. 1989).
The ART--P observations  (2.5--30 keV; Sunyaev et al.  1993;
Markevitch et al. 1993) showed that the diffuse component
follows the distribution of the giant
molecular clouds (GMCs hereafter) present in this region.
These authors  suggested  that  the diffuse emission is due
to scattering in the molecular gas of the X--rays from
nearby X--ray binaries.
They also predicted the presence of a
strong iron line of fluorescent origin at 6.4 keV, that
was  later observed with the  ASCA satellite (Koyama et al. 1996a).
Indeed, the high spectral resolution observations
performed with ASCA
confirmed the presence of the diffuse 6.7~keV line component,
extended symmetrically with respect to the GC,
and discovered the existence
of a diffuse 6.4~keV   line. The latter is spatially
correlated  with the distribution of the molecular clouds and is
particularly intense in the
direction of the   Sgr~B2 molecular cloud (Koyama et al. 1996a).

The  spectrum of the diffuse emission
is well described  by a thermal   hot plasma
with  temperature   $\geq$7~keV, but there is also   evidence
for a multi--temperature, or a non-equilibrium ionization, plasma.
In fact, several emission lines are present, with the K--lines from
iron and sulfur (at $\sim$2.4~keV) particularly bright (Koyama et al. 1996a).

The GC X--ray emission is part of the diffuse  emission that
permeates the Galactic plane
(Kaneda 1997; Valinia \& Marshall 1998), the nature of which   
is still unknown.
Its temperature is too high to allow the confinement
of the emitting plasma
by Galactic gravity (Townes 1989);
part of it
could be due to
thermal emission from supernova remnants (SNRs), or to the integrated
contribution of discrete weak sources (Watson et al. 1981; Zane et al. 1996),
but other emitting processes
have been proposed, such as
non--thermal emission from SNRs, inverse Compton
scattering by relativistic
electrons (Skibo \& Ramaty 1993),
emission lines of non--thermal origin
produced during capture of  electrons by accelerated ions
(Tatischeff et al. 1998),
charge exchange by low energy
heavy ions with neutral gas (Tanaka et al. 1999),
non--thermal emission from the
interaction  of low energy cosmic ray electrons with the interstellar
medium (Valinia et al. 2000).

The 6.4~keV iron line component is thought to be due to irradiation of the
clouds by hard photons
produced by bright X--ray sources (Koyama et al. 1996a, 
Murakami et al. 2000, Murakami et al. 2001),
located inside or outside the cloud (Fromerth et al. 2001).
However, the line emission seems too intense  to be due to
reprocessing of hard X--rays   from
any known  source (external to the cloud) in the GC region.
A possibility for the illuminating source is the GC itself,
the putative massive black-hole Sgr~A* (Ghez et al. 1998),
during a past phase of
high-energy activity (Koyama et al. 1996a;
Churazov et al. 1996).

We report here on new X--ray observations of three
giant molecular clouds of the GC region: Sgr~B, Sgr~C and Sgr~D.
The data were obtained during a survey of the GC region
performed with the BeppoSAX satellite in 1997--1998.
The results on the population of discrete bright sources
have been reported elsewhere (Sidoli et al. 1999; Sidoli 2000), while those
on the diffuse emission from the Sgr~A region
can be found in Sidoli \& Mereghetti (1999).

\section{Molecular clouds in the Galactic Center region}

The interstellar medium (ISM) 
in the inner 500 pc of the Galaxy is   dominated
by molecular gas (G\"usten 1989) mostly contained in giant
molecular clouds. These clouds have
peculiar properties, compared to those of
the Galactic Disk clouds: higher turbulence, higher
densities ($\geq10^{4}$~cm$^{-3}$) and
higher average kinetic temperatures ($T$$\sim$70 K, H\"uttemeister et al. 1993).
High densities are indeed required against tidal disruption
in the gravitational potential  of the GC region (G\"usten \& Downes 1980).
The high kinetic temperatures  are probably produced
by shocks and dissipation of turbulence driven
by the differential Galactic rotation (Wilson et al. 1982).

The ISM in the GC environment has been surveyed in a large variety
of molecules, each tracing  gas with
different densities: $^{12}$CO (Oka et al. 1998),
$^{13}$CO  (Bally et al. 1987; Heiligman 1987),
C$^{18}$O (Dahmen et al. 1998),
CS (Tsuboi et al. 1999),
NH$_{3}$ (Morris et al. 1983), HCN (Jackson et al. 1996),
HNCO and CH$_{3}$CN (Bally et al. 1987),
SiO (Martin--Pintado et al. 1997; H\"uttemeister et al. 1998).

These observations reveal that  the GC  clouds are mostly distributed
at positive Galactic longitudes, display  complex kinematic properties
and have typical dimensions of  $\sim$10 pc
(see Morris \& Serabyn 1996 and Mezger et al. 1996 for the most
recent reviews).
The main molecular clouds
in the GC region are
Sgr~B, Sgr~D and Sgr~C. Their relative positions
are still not well determined and
they are collectively named ``the Galactic Center molecular
clouds complex".
This complex has also been surveyed in the radio
continuum (e.g. Anantharamaiah et al. 1991; Gray 1994; La Rosa et al. 2000)
revealing
both thermal and non--thermal
emission.
The complexity of the observed structures (possibly physically related
with these clouds)
is outlined in the following subsections.

\subsection{Sagittarius B2}

The Sgr~B2 GMC  is located at a projected distance from
the GC of $\sim$120~pc.

Radio continuum observations reveal that Sgr B2 is one of the most active
star-forming regions in our Galaxy. It contains
about 50 very compact H\,{\sc ii} regions, each excited by a newly formed
single O/B massive star (see Gaume et al. 1995
for a complete census of these regions).
The main H\,{\sc ii} regions inside the Sgr~B2 molecular cloud 
are located in the
North--South direction. The principal group of these
compact regions is called
``Sgr~B2 Main complex'' (Sgr~B2(M)).
Also shell--like  and cometary--shaped H\,{\sc ii} regions
are present, maybe
due to  bow shocks produced by
O/B stars with a strong wind  moving supersonically
through the molecular
cloud (Van Buren et al. 1990).

Sgr~B2, like all the other molecular clouds, has a non--uniform
distribution of density and temperature.
Different temperature components
are present, as well
as   clumps with density as high as 
$n$$_{\rm H_{2}}\sim10^{6}-10^{7}$~cm$^{-3}$
and size smaller than 0.5~pc.
The average parameters of the cloud are:
$n$$_{\rm H_{2}}\sim10^{5}$~cm$^{-3}$,  diameter
$\sim$20~pc, and  total mass M$\sim$$(5-10)\times10^{6}\msun$
(Lis \& Goldsmith 1991; Numellin et al. 2000).

The Sgr~B1 H\,{\sc ii} region is located South--West of Sgr~B2.

\subsection{Sagittarius C}

A complex of radio structures is present   within a few arcminutes from the
Sgr~C GMC.
The name Sgr~C indicates both a molecular cloud and a
shell--like H\,{\sc ii} region, located near  the South--West edge of the molecular
cloud and   probably physically related to it (Liszt \& Spiker 1995).
North of this shell a straight non--thermal  filament is present, running
perpendicular to the Galactic plane.
Another radio filament extending parallel to the Galactic plane
seems to start, in projection, from the eastern edge of the H\,{\sc ii} region.

\subsection{Sagittarius D}

Different molecular and radio continuum components are
present in this   region:
a GMC marked by the CO molecule (J=3--2 emission);
a peak in the CS   emission (J=5--4 line); a thermal radio
continuum source, the H\,{\sc ii} region G1.13--0.10; the non--thermal
shell of the G1.05--0.15 SNR.

Observations carried out in the above mentioned molecular lines
seemed to indicate that the molecular core traced by CS is
physically related to the H\,{\sc ii} region, but not with the
molecular cloud traced by the CO line (Lis 1991).

The real location of both the Sgr~D  H\,{\sc ii} region and the SNR
with respect to the GC is still
unclear, although H$_2$CO
observations seem to favor the hypothesis that they are not located at the
GC, but beyond it (Lis 1991, Mehringer et al. 1998).

\section{Observations and data analysis}

The data analysed   here were obtained with
the  MECS instrument
(Boella et al. 1997). The MECS  operates in  the 1.8--10 keV energy range,
providing a
moderate spatial ($\sim$1$'$ full width at half maximum (FWHM))
and
energy resolution
($\sim$8.5\%  FWHM at 6 keV)  over a circular
field of view with a diameter of $56'$.
The log of observations is reported in Table~\ref{tab:log}.

\vspace{0truecm}
\begin{table}[ht]
\caption{BeppoSAX observations summary 
}
\label{tab:log}
\begin{tabular}[c]{lrll}
\hline 
Main & Pointing Direction  &   Observation               & Exposure    \\
Target & $l$, $b$        &     Date      & (ks)   \\
\hline
Sgr~B2 &   $ 0.7$,~$-0.04$  &1997 Sep  3--4   &47.6           \\
Sgr C  &   $359.4$,~$-0.11$  &1997 Sep   16    &14.3            \\
Sgr D  &   $1.1$,~$-0.14 $  &1997 Sep  4--6   &45.5       \\
\hline
\end{tabular}
\vspace{0truecm}
\par\noindent
\end{table}

We have not performed a systematic source detection.

In all our spectral analyses, the
counts were rebinned to oversample the FWHM of the energy resolution by
a factor 3 and to have   a minimum of 20 net counts
per energy channel, in order to allow use of the $\chi^2$ statistics.

The spectra of the diffuse emission have been
corrected with the effective area values
appropriate  for extended sources,
generated with the  {\it effarea}  program
available in the SAX Data Analysis System (SAXDAS).
It convolves a flat
surface brightness distribution  with the
energy and position
dependent vignetting and point spread function.
In case of point--like sources,
the standard response matrices, appropriate for the
extraction regions, have been used.

All the spectra have been corrected for a local background,
accumulated from source--free regions
of the same observation.
All uncertainties are quoted at 90\% confidence level.

 
All fluxes reported in the following sections have been measured
within 2$'$ from the peak emission, except in the case of Sgr~B1
and Sgr~D, where  
more complicated extraction regions enclosed by the radio contours 
have been considered.


\section{Results }

\subsection{Sagittarius~B2}

\begin{figure*}[!ht]
\vskip -1.5truecm
\centerline{\psfig{figure=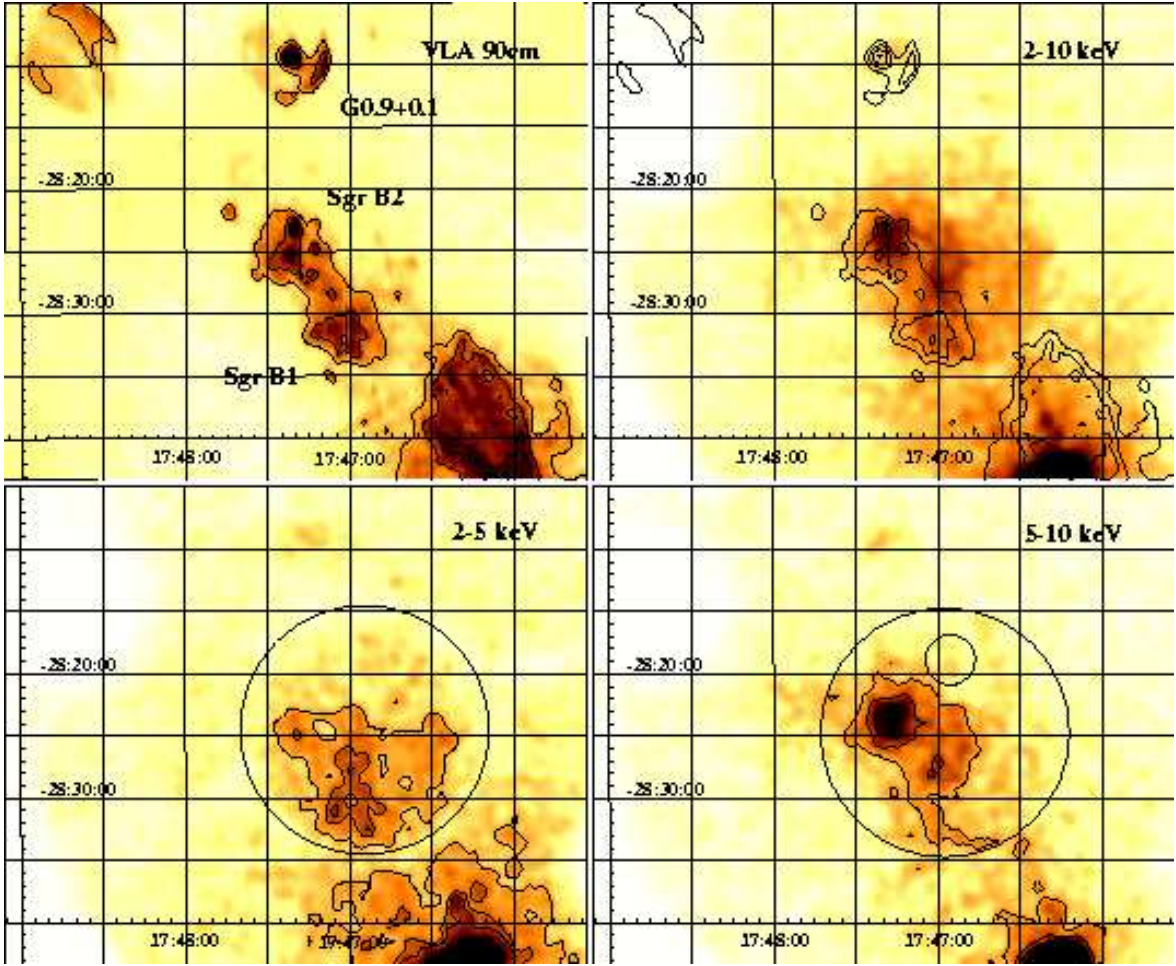,height=140mm,bbllx=70mm,bblly=80mm,bburx=140mm,bbury=250mm
}}
\vskip 1.8truecm
\caption{{\em Upper panels:}
on the left, radio map (VLA, 90~cm; La Rosa et al. 2000)
of the Sgr~B region.
The composite supernova remnant G0.9+0.1 is also visible
in the upper part of the region.
On the right, MECS image in the total band (2--10 keV)
of the same region  with the radio contours overlaid.
{\em Lower panels:} soft (2--5 keV; on the left)  and
hard MECS images (5--10 keV; on the right).
The bright source in the lower right of all the X--ray images is the
low mass X--ray binary 1E1743.1--2843 (Cremonesi et al. 1999).
The smaller circle represents the background extraction region adopted
for the Sgr~B2 spectral analysis. The larger circle marks the position
of the strongback structure of the MECS detector.
All the X--ray images have been smoothed with
a Gaussian with FWHM=1$'$.
The X--ray contour levels represent  95\%, 80\% and 60\% of 
the peak intensity.
The coordinate grids are in the J2000 equinox 
}
\label{sgrb2radiox}
\end{figure*}

The central part of the MECS (2--10 keV)
image pointed on Sgr~B2
is displayed in Fig.~\ref{sgrb2radiox}, together
with a radio map (90~cm)
of the same region obtained with the VLA (La~Rosa et al. 2000).
In the lower panels of the same figure,
the soft (2--5 keV) and hard (5--10 keV)
X--ray images are   shown.
A clear excess, correlated with
the spatial distribution of Sgr~B2, is present. It is more
prominent in the hard X--ray band, while    at softer
energies there is evidence for X--ray emission from Sgr~B1
with a net count rate  of $(1.71\pm{0.11})\times10^{-2}$~s$^{-1}$ 
(2--10~keV).

\begin{figure}[!ht]
\vskip 0truecm
\centerline{\psfig{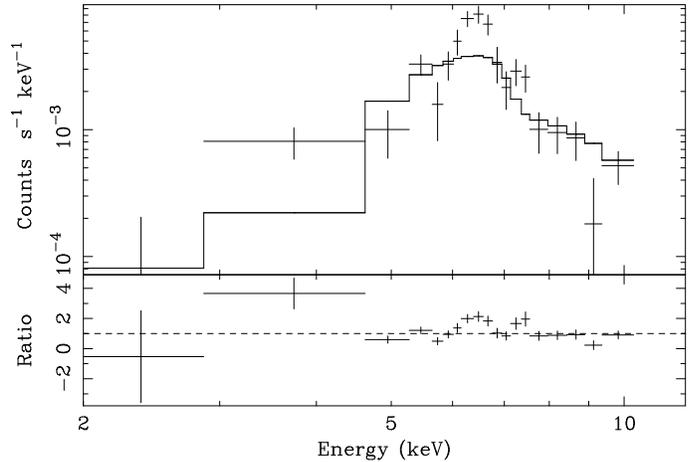}}
\vskip 0 truecm
\caption{MECS counts spectrum of the Sgr~B2 molecular cloud
X--ray emission fitted with a strongly
 absorbed   power-law.  Positive residuals at  $\sim$6--7~keV
are clearly visible, thus requiring the addition of a Gaussian line to the
model  
}
\label{fig:powsgrb2}
\end{figure}

The  peak
of the hard X--ray emission from Sgr~B2 is at  coordinates
R.A.$=17^{{\rm h}}~47^{{\rm m}}~16^{{\rm s}}$,
Dec.$=-28\deg~23'~40''$ (J2000), consistent, within the
position uncertainties ($\sim$1$'$), with that reported by
Murakami et al. (2000; ASCA, 6.2--6.6 keV).
For the spectral analysis we accumulated counts from 
a circle with a 2$'$ radius
centered on this position.
The background extraction region is displayed in Fig.~\ref{sgrb2radiox}.
We verified that also with other backgrounds, always extracted from
source-free regions located inside the strongback structure of the detector,
the spectral analysis gave consistent results.

\begin{figure*}
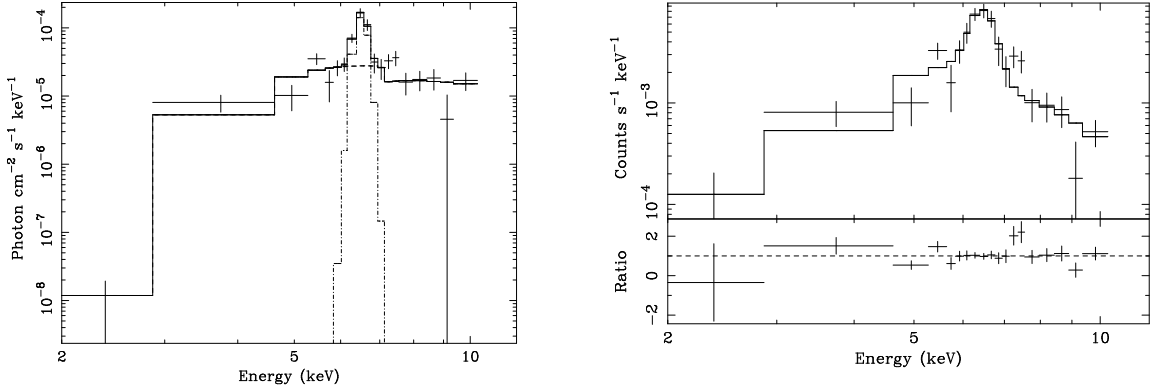

\vspace{0.5cm}
\hbox{\hspace{0.5cm}
\includegraphics[height=6.8cm,angle=-90]{H2651F3a.ps}
\hspace{1.0cm}
\includegraphics[height=7.2cm,angle=-90]{H2651F3b.ps}}
\caption[]{MECS photon spectrum of the Sgr~B2 molecular cloud
X--ray emission (left panel) fitted with  a strongly
absorbed power-law plus a Gaussian line (see the text for details).
The count spectrum is shown in the right panel, together with
the ratio of observed to model counts 
}
\label{fig:powgausgrb2}
\end{figure*}

The fit with a single power-law is unacceptable
($\chi^2$/dof=62.2/16; dof=degrees of freedom). 
It shows both a strong emission line in the
6--7 keV region and a
very high absorption at low energy (see Fig.~\ref{fig:powsgrb2}).
The addition of a  Gaussian line
results in the following best fit parameters:
interstellar absorption 
$N_{\rm H}$=($45 ^{+45} _{-38}$)$\times10^{22}$~cm$^{-2}$,
power-law photon index $\Gamma$=$2.0 ^{+1.5} _{-1.0}$,
line energy $E_{\rm line}$=6.50$\pm0.07$~keV,
line equivalent width EW=($2.0 ^{+1.1} _{-0.8}$)~keV, line
intensity $I_{\rm line}$=$(0.7-1.8)\times10^{-4}$ photons~cm$^{-2}$~s$^{-1}$
($\chi^2$/dof=25.1/13).
The 2--10 keV (or 4--10 keV)  observed flux
is $\sim$1.8$\times10^{-12}$~erg~cm$^{-2}$~s$^{-1}$.
The 2--10~keV flux (corrected for absorption) is
$F_{\rm X}$=6.9$\times10^{-12}$~erg~cm$^{-2}$~s$^{-1}$, which corresponds to
an X--ray luminosity $L$=5.6$\times10^{34}$~erg~s$^{-1}$, for a
distance of 8.5~kpc.
The fit to the spectrum is shown in Fig.~\ref{fig:powgausgrb2}.
There are some positive residuals at 7.4~keV,
which are probably caused by an inappropriate modeling of the continuum
and/or  of the nearby Fe~I edge at $\sim$7.1~keV.
In fact   the addition of a second
Gaussian line is not statistically significant.

Although the fit residuals and $\chi^2$ values  suggest that more
complex models are probably needed,
we performed the following further fits to
the data,
to allow a comparison with the ASCA results (Murakami et al. 2000).

A  power-law plus a  Gaussian line with
fixed  $N_{\rm H}$ (from 10 to 85$\times10^{22}$~cm$^{-2}$)  resulted in
a photon index ranging from 0.17 to 3.3.
The line center was always found at
$\sim$6.5~keV. The line width varied between
$\sigma$=240~eV and $\sigma$=55~eV
and its intensity in the range
$\sim$$(0.9-1.7)\times10^{-4}$~photons~cm$^{-2}$~s$^{-1}$,
corresponding respectively to EWs of 1 and 4.5~keV.
A fit with a free column density and
fixed  photon index $\Gamma$=2, resulted in
$N_{\rm H}\sim(40\pm{15})\times10^{22}$~cm$^{-2}$,
$E_{\rm line}$=(6.50$\pm{0.07}$)~keV,
$I_{\rm line}\sim$1.1$\times10^{-4}$ photons~cm$^{-2}$~s$^{-1}$,
$\sigma$$\sim$$150$~eV, and an EW of $2.2\pm{0.7}$~keV.
The 2--10~keV and 4--10~keV fluxes
corrected for the absorption are
7$\times10^{-12}$~erg~cm$^{-2}$~s$^{-1}$ and
4.5$\times10^{-12}$~erg~cm$^{-2}$~s$^{-1}$, corresponding  to
$L$$\sim5.7\times10^{34}$~erg~s$^{-1}$ and $L$$\sim3.7\times10^{34}$~erg~s$^{-1}$ 
respectively.

These results differ from the  ASCA ones in the absorbing column density
(our value is significantly lower) and in the
energy of the line, which is
significantly higher  compared to the 90\% confidence range
of 6.35--6.45~keV found with ASCA (Murakami et al. 2000).

We tried also a fit  with a power-law plus
two Gaussian lines ($\sigma$ fixed at 0),
with energies fixed at 6.4 and 6.7 keV. This fit ($\chi^2$/dof=24.7/14)
gave
$N_{\rm H}=(30^{+40}_{-20})\times10^{22}$~cm$^{-2}$ and
$\Gamma$=$1.40^{+1.70}_{-1.25}$.
The EWs for the two iron lines,
EW$_{6.4}$=$660^{+670}_{-210}$ and EW$_{6.7}$=$270^{+215}_{-180}$,
are again significantly different from the Murakami et al. (2000) results.
Indeed they found that the flux contributed by
the 6.7~keV is less than 10\% of that from
the 6.4~keV line, and that the profile of the line they
observe is reproduced by the 6.4~keV line alone.
This discrepancy can probably be explained by
the choice for the local background.
Murakami et al. (2000) subtracted from the
main peak also emission coming from the molecular cloud itself and from
the Sgr~B1 region (see their Fig.~1),
while our choice avoids the region overlapping with
the radio contours of both Sgr~B2 and Sgr~B1.

The final comparison with the ASCA results
was performed by extracting the background 
from the same position used in Murakami et al. (2000).
Using their model, an absorbed power-law with the photon index 
fixed at 2 plus a Gaussian line,
we obtained the following results:
$N_{\rm H}=(73^{+35}_{-25})\times10^{22}$~cm$^{-2}$,
line energy of $6.50^{+0.08}_{-0.09}$~keV, 
EW=$1.96^{+2.04}_{-0.77}$~keV ($\chi^2$/dof=20.1/12).
Thus, in this case, the absorbing column density is compatible
with ASCA results. The energy of the iron line is now
compatible with the SIS results, but still higher than the GIS
90\% confidence range.

 
We finally tried with another reasonable
model for a molecular cloud emitting X--rays: a high temperature plasma 
model ({\sc mekal}). 
The fit is unacceptable ($\chi^2$/dof=43.8/16), 
resulting in positive residuals around 6.3~keV 
($N_{\rm H}$$\sim$$6\times10^{23}$~cm$^{-2}$; kT$\sim$4~keV).
The addition of a Gaussian line  ($\chi^2$/dof=25.2/14)  
at 6.45~keV (6.30--6.55 keV 
90\% confidence range) resulted in an EW of $\sim$1~keV and in
a temperature kT$\geq$6~keV with an absorbing column
$N_{\rm H}$$\sim$$4\times10^{23}$~cm$^{-2}$.

\subsection{Sagittarius~C}
 
\begin{figure*}[!ht]
\vskip -3.0truecm
\centerline{\psfig{figure=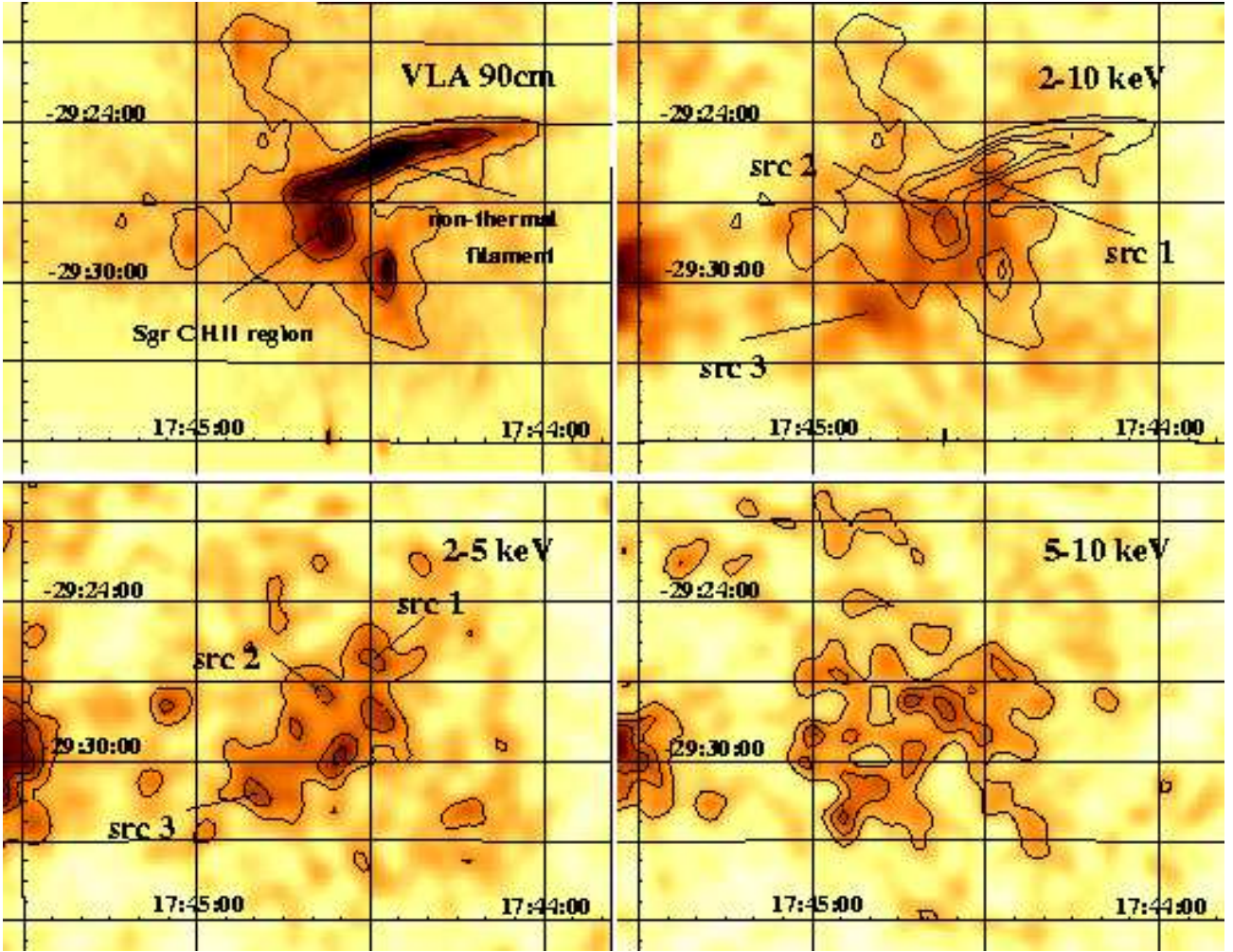,height=155mm,bbllx=70mm,bblly=80mm,bburx=140mm,bbury=250mm}}\vskip 1.6truecm
\caption{{\em Upper panels:} on the left,
radio map
(VLA, 90~cm; La Rosa et al. 2000) of the Sgr~C region.
The central blob of radio emission is the Sgr~C H\,{\sc ii} region,
while the straight structure is a   non--thermal filament running
perpendicular to the Galactic plane.
On the right, the same region of the sky, imaged with the
MECS (2--10 keV) instrument. The radio
contours are overlaid  for comparison.
{\em Lower panels:}  soft   (2--5 keV; on the left)  and
hard MECS images (5-10 keV; on the right).
All the X--ray images have been smoothed with
a Gaussian with FWHM=1$'$.
The X--ray contour levels represent  95\%, 80\% and 60\% of 
the peak intensity.
The coordinate grids are in the J2000 equinox 
}
\label{sgrcradiox}
\end{figure*}

The X--ray and radio images of
Sgr~C are compared in Fig.~\ref{sgrcradiox}.
The  soft and hard X--ray images (lower panels of Fig.~\ref{sgrcradiox}) are quite similar.
A North--West to South--East
excess, with three main peaks, is visible in the total X--ray band.
The Northern  peak (src~1), located at
R.A.$=17^{{\rm h}}~44^{{\rm m}}~28^{{\rm s}}$,
Dec.$=-29\deg~26'~15''$ (J2000, $\sim1'$ error), could be
associated
with a hot spot in the
non--thermal radio filament that ends at the Northern  edge of Sgr~C
(Fig.~\ref{sgrcradiox}).
The central X--ray  peak (src~2) is
positionally coincident with
the Sgr~C H\,{\sc ii} region (R.A.$=17^{{\rm h}}~44^{{\rm m}}~38^{{\rm s}}$,
Dec.$=-29\deg~27'~44''$ (J2000)), while the
South--Eastern peak (src~3, at R.A.$=17^{{\rm h}}~44^{{\rm m}}~50^{{\rm s}}$,
Dec.$=-29\deg~31'~01''$ (J2000)) has no counterparts in the VLA radio map.

Their net count rates in the 2--10~keV energy range are
the following: $(6.28\pm{1.67})\times10^{-3}$~s$^{-1}$ (src~1),
$(1.06\pm{0.18})\times10^{-2}$~s$^{-1}$  (src~2, Sgr~C H\,{\sc ii} region)
and $(8.94\pm{1.75})\times10^{-3}$~s$^{-1}$ (src~3).

\subsection{Sagittarius~D}

\begin{figure*}[!ht]
\vskip -3.0truecm
\centerline{\psfig{figure=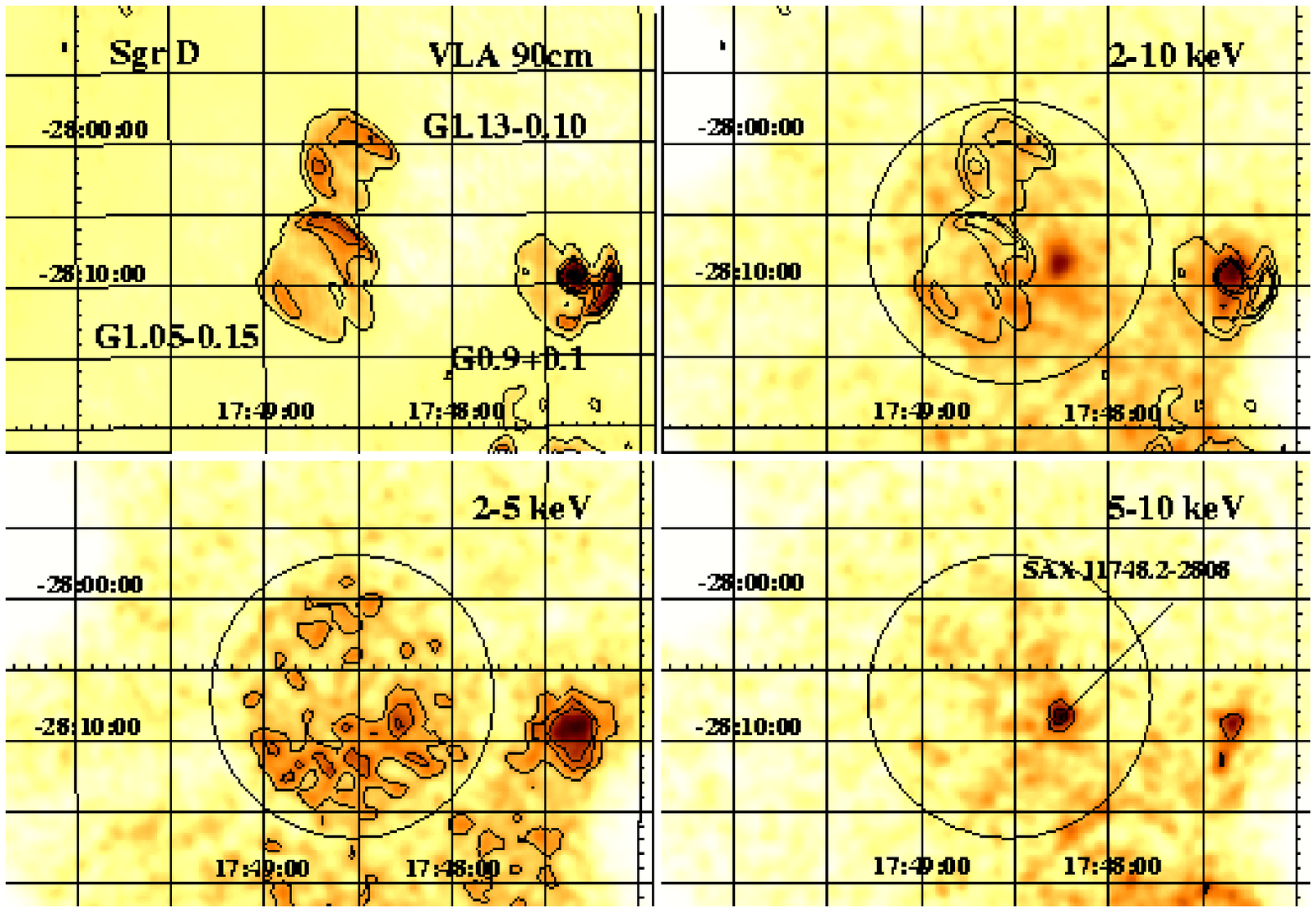,height=150mm,bbllx=70mm,bblly=80mm,bburx=140mm,bbury=250mm}}
\vskip 0.7truecm
\caption{{\em Upper panels:}
on the left,
radio map (VLA, 90~cm; La Rosa et al. 2000) of the Sgr~D region.
On the right, the same region of the sky imaged with
the MECS (2--10 keV) instrument.
{\em Lower panels:} X--ray images, in two energy ranges: soft (2--5 keV)
on the left, hard (5--10 keV) on the right.
The big circle in the three MECS images represents 
the position of the circular
detector structure of the strongback.
All the X--ray images have been smoothed with
a Gaussian with FWHM=1$'$.
The X--ray contour levels represent  95\%, 80\% and 60\% of 
the peak intensity.
The coordinate grids are in the J2000 equinox 
}
\label{sgrdradiox}
\end{figure*}

Two well defined and contiguous
radio shells are evident in the VLA observation  
(La~Rosa et al. 2000) of Sgr~D
shown in Fig.~\ref{sgrdradiox}.
The Northern  shell is the
Sgr~D H\,{\sc ii} region (G1.13--0.10), while the Southern one
is the   supernova remnant G1.05--0.15.
The comparison with the MECS image shows
weak X--ray emission spatially correlated
with some parts of the radio shells,
in particular with the Southern rim of the SNR G1.05--0.15 (especially
at soft X--ray energies).
The MECS count rate from this diffuse structure
is $(5.10\pm{0.79})\times10^{-3}$~s$^{-1}$ (2--10~keV).
If this emission is really associated with G1.05--0.15, this is
the first detection of this SNR  at X--rays.
We note that the emission is statistically significant, but its
curved shape is probably
an instrumental artifact due to partial absorption
caused by the support structure of the detector window, a circular
rib (marked by the big circle in Fig.~\ref{sgrdradiox}) located at
$\sim$10$'$ from the  center of the MECS field of view.

Low surface brightness X--ray emission is also present
in the direction of the Sgr~D
H\,{\sc ii} region,
with a net count rate of
$(8.1\pm{1.1})\times10^{-3}$~s$^{-1}$ (2--10~keV).

\begin{figure}[!ht]
\vskip 0.4truecm
\centerline{\psfig{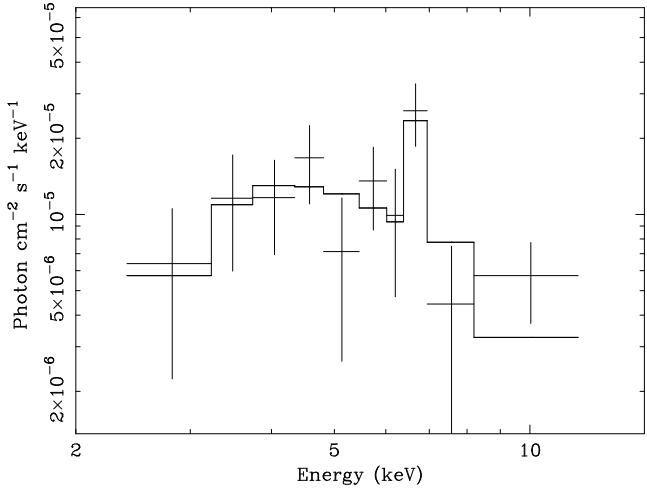}}
\vskip 0.3truecm
\caption{Observed MECS photon spectrum of SAX~J1748.2--2808, together
with a {\sc mekal} fit (histogram) 
}
\label{fig:sgrdfe_spe}
\end{figure}
\subsubsection{A new source: SAX~J1748.2--2808}
A point--like X--ray
source, without any radio counterpart at 90~cm (see Fig.~\ref{sgrdradiox}),
is   visible (particularly in the 5--10 keV image) at
R.A.$=17^{{\rm h}}~48^{{\rm m}}~16^{{\rm s}}$,
Dec.$=-28\deg~08'~13''$ (J2000, $\sim1'$ error).
This newly discovered source, that we designate SAX~J1748.2--2808,
has a rather hard and/or strongly absorbed spectrum.
The extraction of counts within 2$'$ from  its position
yields a net count rate in the 2--10 keV range of
$4.84\pm{0.66}\times10^{-3}$~s$^{-1}$.
A strong iron line appears to be present 
in the spectrum of SAX~J1748.2--2808.
The fit with
a power-law plus a Gaussian line
led to the following best fit parameters ($\chi^2$/dof=3.4/5):
$N_{\rm H}$=($12^{+15} _{-12}$)$\times10^{22}$~cm$^{-2}$,
$\Gamma$=1.9$^{+1.7} _{-1.8}$,  $E_{\rm line}$=$6.62\pm0.30$~keV,
EW=$1.2^{+1.0} _{-0.5}$~keV.
The 2--10 keV flux corrected for the absorption is
$F_{\rm X}$=$1.28(^{+1.69} _{-0.54})\times10^{-12}$~erg~cm$^{-2}$~s$^{-1}$, 
corresponding to
a 2--10 keV luminosity $L$=$10^{34}$~erg~s$^{-1}$
(assuming d=8.5~kpc). 
Taking into account the
uncertainty  introduced by the poorly constrained spectral parameters,
the source luminosity with this model can range from
$6\times10^{33}$ to $2.4\times10^{34}$ erg~s$^{-1}$ (at 8.5~kpc).
 
An equally good fit ($\chi^2$/dof=4.9/7) was obtained with a thermal
equilibrium plasma
model ({\sc mekal} in {\sc xspec}) with solar abundance,
with the following results:
$N_{\rm H}$=($13^{+13} _{-6}$)$\times10^{22}$~cm$^{-2}$,
kT=$5.8^{+35} _{-3.7}$~keV,
$F_{\rm X}$$\sim$$1.36\times10^{-12}$~erg~cm$^{-2}$~s$^{-1}$.
The emission measure derived from the normalization of the {\sc mekal}
model is $n_{\rm e}^{2}$$V$$\sim$1.3$\times$10$^{57}d_{10}^2$~cm$^{-3}$,
where $n_{\rm e}$ is the electron density,  $V$ the emitting volume and
$d_{10}$  the source distance in units of 10~kpc. 
The photon spectrum is displayed in Fig.~\ref{fig:sgrdfe_spe}.
The absorbing column density in this model 
translates into an optical extinction 
$A_{\rm V}$$\geq34$~mag, with a 
preferred value of $A_{\rm V}$=73~mag (Predehl \& Schmitt 1995).

Acceptable fits were also obtained with
blackbody ($\chi^2$/dof=4.0/5) or
bremsstrahlung  models ($\chi^2$/dof=3.5/5)
with the addition
of a Gaussian line at $\sim$6.6~keV.
In conclusion, the poor statistics does not allow us to 
discriminate between
thermal and non--thermal models for the X--ray emission
from SAX~J1748.2--2808, even if
the presence of an iron line  at 6.6~keV favors the
thermal hypothesis.

SAX~J1748.2--2808 has also been serendipitously
detected at large off-axis angle 
during a MECS observation
of the composite SNR  G0.9+0.1
(Mereghetti et al. 1998; Sidoli et al. 2000),
performed on 1999 August 25--27.
Using this observation (net exposure time of 78.5 ks), we
derived for SAX~J1748.2--2808
a net count rate of $(5.1\pm{0.4})\times10^{-3}$~s$^{-1}$
(2--10 keV, corrected for the vignetting). 
This is compatible with
the count rate obtained during the Sgr~D observation, 
carried out two years earlier.
Thus, SAX~J1748.2--2808 does not show evidence for long--term variability.
The large off-axis angle prevented us from performing 
a detailed spectral analysis. 
All we can say is that, using a power-law plus a Gaussian 
model and fixing the column density and the power-law index to 
the best fit results reported above, the 90\% confidence ranges
for the iron line centroid and the EW are: 
6.2--7.7 keV and 0.04--1.3~keV.
Thus,   the iron line emission does not show evidence for
variability between the two observations either.

A search in the ASCA public archive resulted in only 
one observation
of the SAX~J1748.2--2808 field, performed on 1996 September 19,
with an exposure time of $\sim$22.3 ks.
No source can be detected at the SAX~J1748.2--2808 position, with
a 3$\sigma$ upper limit of 1.34$\times10^{-2}$~s$^{-1}$ 
(GIS; corrected for the vignetting). 
This translates into 
an unabsorbed flux, $F_{\rm X}$, of
$2.22\times10^{-12}$~erg~cm$^{-2}$~s$^{-1}$ 
(power-law best-fit model; 2--10 keV),
a factor of 1.7 larger than our MECS detection, 
and thus   still compatible with the lack of
 the long-term variability.

\section{Discussion}
 
During the observations of three fields centered on giant
molecular clouds in the GC region, we detected X--ray emission
from several sources of different kind, both diffuse and point-like.
None of them have ROSAT X--ray counterparts 
(Sidoli et al. 2001).

\subsection{A newly discovered point-like source: SAX~J1748.2--2808}
\label{sect:sgrdfe}
 
\begin{figure}[!ht]
\vskip -1.6truecm
\centerline{\psfig{figure=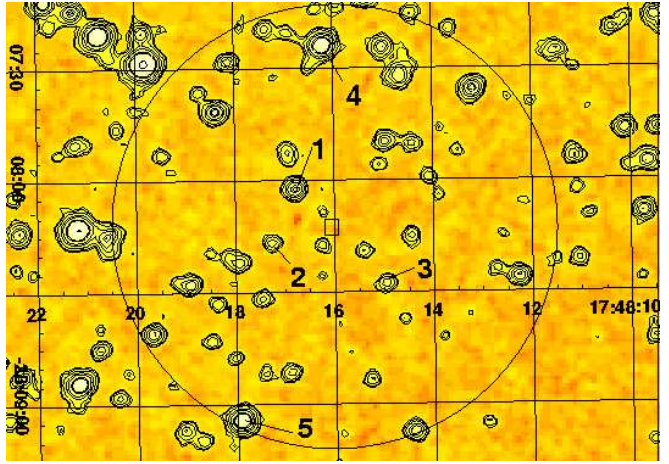,height=78mm,bbllx=70mm,bblly=80mm,bburx=140mm,bbury=250mm}}
\vskip 1truecm
\caption{Optical (red band) image of the uncertainty 
region of SAX~J1748.2--2808.
The data are from the ``Second Epoch Survey  of the Southern Sky''
made by the Anglo--Australian Observatory with the
UK Schmidt Telescope and provided by the ESO Online Digitized Sky.
The large circle marks the MECS error box (1$'$ radius), the center
of which is indicated by the square.
The numbers 1, 2 and 3 
indicate the three nearest catalogued stars (0$\farcm$255, 0$\farcm$262
and 0$\farcm$330 offset from the X--ray source, respectively).
They have  the
following R and B magnitudes: 16.5 and 18.9 (star n.1), 17.4 and 20.3
(star n.2), 17.4 and 20.1 (star n.3).
Numbers 4 and 5 mark the two brightest (in R magnitude) stars inside the
X--ray error box; they have  R and B magnitudes of 15.2 and
17.2 (star n.4), 15.3 and 18.0 (star n.5)}
\label{fig:opt}
\end{figure}

The optical image of the position of SAX~J1748.2--2808, derived from the
digitized sky survey
provided by ESO/ST-ECF Science Archive
is shown in  Fig.~\ref{fig:opt}.
Eighteen catalogued stars,
with R magnitude in the range 15.2--17.4 (and B$\sim$17.2--20.4),
are located
inside the X--ray error circle of SAX~J1748.2--2808 (1$'$ radius).
The derived optical to X--ray flux ratio 
log($f_{\rm X}/f_{\rm opt}$) is always $>$$-1$, 
thus making  an association of SAX~J1748.2--2808 with
one of these stars unlikely (Maccacaro et al. 1988).

A search in the SIMBAD database resulted in two possible
counterparts:
the infrared source IRAS~17450--2807
and the SiO maser source SiO~17450--2808
(10$''$ off-set from the IRAS source),
first reported as ``object number~6'' in the catalogue of
SiO maser sources by Shiki et al. (1997).
Masers of stellar origin can be excited by radiative or collisional
pumping in the outer atmosphere of red (super)giants
(Matsuura et al. 2000),
around Asymptotic Giant Branch (AGB) stars with large mass loss
(Desmurs et al. 2000), or around young stellar objects, where the
outflow of material interacts with the surrounding medium
(Shepherd \& Kurtz 1999).

The SiO maser radial velocity $V_{\rm LSR}$ of +101.2~km~s$^{-1}$
is  appropriate for a source  located in the GC region, thus suggesting
a possible association with the Sgr~B2 molecular cloud
(Shiki et al. 1997).
The colors of the infrared source IRAS~17450--2807 make
it a likely young stellar object,
even if the possibility of a late--type star
cannot be completely excluded (Shiki et al. 1997).
The IRAS source variability
suggests  that it could be a Mira--type variable star.
A  search in the literature resulted in a further  maser source
coincident (within the positional uncertainties) with SAX~J1748.2--2808:
a 1667 MHz OH maser source
(named ``E'' in Mehringer et al. 1998)
with a velocity of $V_{\rm LSR}$=+75~km~s$^{-1}$.
Mehringer et al. (1998) favor the hypothesis that the maser source
is associated with an evolved star
since it is not spatially correlated with any known H\,{\sc ii} region.
We note that the relatively large uncertainties in their   coordinates
are consistent with the possibility that these
four sources  (SiO and OH masers, IRAS source and
SAX~J1748.2--2808)  are the same object.

If the association with the SiO maser source is real, this places the
X--ray source at the same approximate distance as the Galactic center,
and thus it will have an X--ray luminosity of approx. $10^{34}$ erg
s$^{-1}$, with an apparently quiescent emission and a rather high
temperature of approx. 6~keV. 
Young stellar objects, and in particular
protostars, indeed often have high coronal temperatures, of $\geq 5$
keV, even in their non-flaring emission (e.g. as reported by Ozawa et
al. 1999 for protostars in Orion), and thus the hypothesis of it being
a single protostar would be compatible with the observed
spectrum. However, the X--ray luminosity is, specially for a
non-flaring source, too high for an individual protostar. While
quiescent coronal emission has only been observed in a limited number
of protostars, the observed X--ray luminosities seems to be correlated
with their mass (see Feigelson \& Montmerle 1999), with luminosities
limited to few times $10^{31}$ erg s$^{-1}$ (e.g. Ozawa et al. 1999)
at the ``low-mass'' end (i.e. around a solar mass) ranging to up to
$\ge 10^{32}$ erg s$^{-1}$ for more massive protostars (e.g. Nakano et
al. 2000). 

The X--ray luminosity of SAX~J1748.2$-2808$ is therefore higher than
the one of protostars observed up to now. If it indeed the emission is
protostellar in origin, the two possibilities are that either it is a
protostar more massive than observed up to now in the X--rays, and thus
more luminous (assuming that the mass-luminosity correlation extends
to higher masses than observed thus far) or that the emission is the
result of the superposition of a number (10 or so) high-mass
protostars of the type observed by Nakano et al. (2000) in NGC~2264
and which form a compact (and unresolved) cluster. 
The second
hypothesis would be compatible with the observed lack of temporal
variability, with the flaring behavior of individual sources likely
time-averaged in the ensemble to yield a smooth light curve.

The other possibility is that SAX~J1748.2$-2808$ is the X--ray emission
from a giant molecular cloud core (or compact H\,{\sc ii}
region). 
Emission from H\,{\sc ii} region cores has been observed with
characteristics very similar to the ones of SAX~J1748.2$-2808$. 
For
example, Hofner et al. (1997) have observed, with ASCA, the core of
the W3 region complex, finding a compact (barely resolved) source with
an X--ray luminosity of $\simeq 2 \times 10^{33}$ erg s$^{-1}$. 
The
spectrum shows a visible Fe~K line, and it is compatible with being due
to thermal emission from a plasma with a temperature of $\simeq 6$
keV. 
In addition to the possibility of emission from a collection of
protostars, Hofner et al. (1997) discuss the possibility that the
X--ray emission comes from the hot, wind-shocked cavity which is
produced by the interaction of the strong stellar wind coming from the
more massive stars and the surrounding dense molecular gas. 
The
observed temperature requires wind speeds of about 2000~km~s$^{-1}$,
compatible with the wind speeds from massive stars, and the observed
X--ray luminosity is compatible with the expected cavity size. 
This
model would thus well explain the emission from SAX~J1748.2$-2808$,
and would naturally result in a constant X--ray emission with no
flaring activity (as would be common for a protostellar source). 
More
recently, molecular cores with similar X--ray emission, both in terms
of temperature and luminosity, have been observed with ASCA in the 
NGC~6334 giant molecular cloud (Sekimoto et al. 2000). 

If indeed the emission from SAX~J1748.2$-2808$ is due to a shocked
molecular core cavity, the strong Fe~K line observed is apparently more
intense than observed in e.g. the W3 core. 
This could have an impact
on the expected contribution of the emission from molecular cores on
the observed Fe~K line emission in the Galactic ridge 
(Yamauchi \& Koyama 1993). 
On the basis of the relatively weak Fe~K emission
observed in the W3 core Hofner et al. (1997) consider that the
contribution of molecular cores to the total Fe~K ridge emission is
likely to not exceed $\simeq 2\%$ of the total. 
If however the
observed stronger Fe~K line of SAX~J1748.2$-2808$ is indeed
representative of such cores, a significant fraction of the total
Fe~K ridge emission could effectively be due to them. 
 
We have inspected the maps of the GC region in different
molecular lines (e.g., Tsuboi et al. 1999), 
to look for possible spatial correlation with
the position of SAX~J1748.2$-2808$,
without finding any convincing association.
 
Another possibility is that of a nearby stellar object with a strong
intrinsic absorption ($N_{\rm H}$$\sim$10$^{23}$~cm$^{-2}$, that
is common among protostars).
However this, if the association with the maser sources is real,
is in contrast with their high velocity that  favors an
object located in the GC region.
 
The X--ray luminosity of SAX~J1748.2--2808 (if the assumed distance, also
indicated by the high interstellar absorption,  is correct)
makes an association with an X--ray binary also possible,
although the
intense iron line emission is   unusual for Low Mass X--ray 
Binaries, where the EW is $<$200 eV, indipendent 
from the source luminosity  (see e.g. Asai et al. 2000).  
A different possibility could be a High Mass X--ray Binary (HMXB).
A large fraction of HMXBs are accreting pulsars (e.g. van Paradijs 1995).
In these systems  rather
large and highly variable EWs (up to 1.8~keV) 
for the fluorescent Fe~K line 
have been observed  (e.g. GX301--2, Leahy et al. 1989).  
However,  flux modulations on the
spin and orbital period should be detected as well. 
Although a spin modulation cannot be ruled out due to the poor statistics,
the absence of long-term flux variability does not favor the
possibility that SAX~J1748.2--2808 is an X--ray pulsar.
However, the huge interstellar absorption ($>$34 mag) 
prevent us from excluding the presence of
a high mass companion star.

The possibility that  SAX~J1748.2--2808 is a cataclysmic variable
should also be considered.
X--ray emission from white dwarves can usually be fitted with thermal
bremsstrahlung models with
temperatures in the range 1--5~keV (Cordova 1995), but
also harder spectra are observed (Yoshida et al. 1992).  They have
luminosity up to 10$^{33}$~erg~s$^{-1}$ and K$_{\alpha}$ iron lines.
Both the energy of the Fe line and the temperature of the spectrum
match this hypothesis. The observed flux would in this case indicate a
distance of only 3~kpc,   which is difficult to reconcile with
the high absorption.

The possible association with a background AGN is also unlikely,
due to the large EW of the iron line 
(typical ranges for Fe~K lines EWs are 100--500 eV; see e.g. Nandra 2000). 
 
In order to explore the possibility of its association with the Sgr~D 
SNR,
we calculated the lower limit $t_{\rm travel}$ to the time required by the
neutron star to travel from the center of Sgr~D SNR to its present
position, about 5--6$'$ away.
This can be expressed by
$t_{\rm travel}=1.4\times10^{5}\times(d_{10}/v_2)$~yr,
where $d_{10}$ is the source distance expressed in units of 10 kpc,
and $v_2$ is the neutron star velocity, in units of 100~km~s$^{-1}$.
We can compare this time with the age  $t_{\rm shell}$
of the Sgr~D SNR as derived from a Sedov
solution, using the dimensions of the radio shell (radius of about 4--5$'$).
In this framework,
$t_{\rm shell}$=11,000$\times(n_{\rm ISM}/E_{\rm 51})^{1/2}\times(d_{10})^{5/2}$~yr
(using a 5$'$ shell radius),
where $n_{\rm ISM}$ is the interstellar medium density in units of cm$^{-3}$,
and $E_{51}$ is the energy of the supernova explosion, in units of
$10^{51}$~erg.
If we reasonably
assume $d_{10}$=1,
$v_2$=1, $n_{\rm ISM}/E_{51}$=1,  then we get
$t_{\rm travel}=1.4\times10^{5}$~yr for the neutron star travel time,
and  about $t_{\rm shell}$=11,000~yr for the adiabatic expansion
of the SNR shell, an order of magnitude difference.
Alternatively, equating these two times
($t_{\rm travel}$$\sim$$t_{\rm shell}$), and assuming
the Galactic Center distance and $E_{51}$=1, we need a very
high velocity for the neutron star
(velocity=1400~km~s$^{-1}$; assuming
$n_{\rm ISM}$=1) or a high density environment
($n_{\rm ISM}$=200, assuming $v_2$=1),
or, more realistically, $v_2$ in the range 2--4
(200--400~km~s$^{-1}$) and
$n_{\rm ISM}$$\sim$$15-50$~cm$^{-3}$.
Thus, a physical connection with Sgr~D SNR is a possibility, but
up to now 
no isolated neutron stars have been observed
to show  such an intense Fe~K line emission.
The observability of X--ray emission lines from the surface of 
magnetized neutron stars seems indeed to be unlikely (Yahel 1982).

In conclusion, we favor the hypothesis that  the SAX~J1748.2--2808 
X--ray emission is produced by protostars (a collection of them
or a single high luminosity object) located at 
the GC distance, 
or by  a  giant molecular cloud core (or compact H\,{\sc ii}
region) shocked  by the  strong stellar wind coming from the
more massive stars.

\subsection{X--ray emission from molecular clouds}
 
Molecular clouds can emit X--rays in different ways.
In several cases the emission is
produced in the star-forming regions naturally located
inside the clouds.
Pre--main sequence stars are   strong  X--ray emitters
(up to 10$^{30}-10^{31}$ erg~s$^{-1}$) with a large variety of behaviors,
showing
both  persistent thermal emission and hard flares 
(e.g. Koyama et al. 1996b).
The $Ginga$ satellite provided the first evidence 
that pre--main sequence stars
emit X--rays above 4~keV. Observing the
$\rho$ Ophiuchi cloud star-forming region, rich in T~Tauri stars
and embedded infrared sources, Koyama et al. (1992) 
obtained a best fit spectrum with
a thermal plasma model with kT$\sim$4.1~keV and an
emission line from He-like iron at 6.6~keV.
The ASCA satellite revealed that T~Tauri stars, and embedded
infrared sources with no optical counterparts,
can emit X--rays also above 8~keV, with  luminosities
ranging from 10$^{29}$~erg~s$^{-1}$ to 10$^{31}$~erg~s$^{-1}$
(reaching 10$^{32}$~erg~s$^{-1}$ during flares).
These observations also confirmed the important fact that hard X--ray
emission can be produced by pre--main sequence
stars  not only  during flares (Koyama et al. 1994).

In the previous section we discussed
the nature of a  newly discovered X--ray source,
finding that its  association with several
young stellar objects is a  likely possibility.
It is possible that also the emission from the molecular
core of Sgr~B2 might be due
to several (5--6) objects
of the same kind of SAX~J1748.2--2808.

Another possibility is the emission from old isolated neutron
stars (ONS) accreting from the dense ISM inside the molecular
cloud  (see e.g., Treves et al. 2000).
The possibility that the integrated   luminosity  from many ONSs might be,
at least in part,
responsible for the   diffuse X--ray emission from the
GC   region was first suggested by Zane et al. (1996; 
see also Pfahl \& Rappaport 2000 for a
recent investigation of X--ray emission from accreting ONSs in
globular clusters). 
Here we follow their approach in evaluating the
expected flux produced by accrecting ONSs inside the Sgr~B2 cloud.
The X--ray luminosity contributed by a single ONS depends on the
density $n_{\rm cloud}$ of the molecular cloud and on the relative
velocity v of the neutron star with respect to the accreting
matter: $L_{\rm ONS}\sim0.7\times10^{33}$~($n_{\rm
cloud}/10^4$)~(v/100)$^{-3}$~erg~s$^{-1}$.
We assume that the
spatial and velocity distributions of ONSs follow that of
low-mass stars near the GC:
$n_{\rm ONS}\sim 4\times10^{3}~{\rm r}^{-1.8}$~pc$^{-3}$ and
f(v)~$\propto ({\rm v}^2/\sigma_{\rm v}^3)\exp{-(3{\rm v}^2/2
\sigma_{\rm v}^2)}$, where r is the distance from the GC, and
$\sigma_{\rm v}$  is the
velocity dispersion.
The monochromatic
flux is given by F$_\nu \sim$~V$_{\rm cloud}$~$n_{\rm ONS}$~$\int
{\rm f(v)}({\rm L}_\nu/4\pi{\rm D}^2){\rm A}_\nu\, d{\rm v}\, ,$
where V$_{\rm cloud}$ is the cloud volume, A$_\nu$ the interstellar
absorption and $L_\nu$ the monochromatic luminosity of the single
source. The latter is taken to coincide with that produced by a
blackbody emitter at the star effective temperature,
T$_{\rm eff}\sim
3(L_{\rm ONS}/10^{33})^{1/4}{\rm f}^{-1/4}$~keV,
where f is the fraction
of the star surface covered by accretion. For the relevant values
of the parameters, $n_{\rm cloud}\sim 10^{5}$~cm$^{-3}$,
$\sigma_{\rm v}\sim 75$~km~s$^{-1}$, cloud radius $\sim 10$ pc,
the spectrum peaks around  $\sim 5$ keV, assuming
f=0.01, with a total flux of
about $4\times 10^{-12}$~erg~cm$^{-2}$~s$^{-1}$,
close to the observed one.
Larger values of f result in softer spectra which are much more severely
absorbed ($N_{\rm H}\sim 10^{24}$~cm$^{-2}$ in the present case) and fail to produce
the observed flux. The total number of ONSs in the cloud is $\sim 200$,
and the typical luminosity for a single source is $\sim 2\times
10^{34}$~erg~s$^{-1}$ which gives an absorbed flux of $\sim 2\times
10^{-14}$~erg~cm$^{-2}$~s$^{-1}$.
Despite the very large ISM density the Thomson depth
within the cloud is $\tau_{\rm T}\approxlt 1$, so the emission does not come
from a photospheric region.  
The presence of a large number of discrete
sources of hard photons is not in contrast with the finding
that most of the interstellar gas is neutral.
The Str\"omgren radius, in fact,
is given by R$_{\rm S} \sim 3\times 10^{15}
(n_{\rm cloud}/10^4)^{-5/12}({\rm v}/100)^{-3/4}{\rm f}^{1/12}$~cm
(see Blaes et al. 1995 for a thorough
discussion), much smaller than the average ONS separation,
$\sim {\rm n}_{\rm{ONS}}^{-1/3}\sim 10^{18}$~cm.

Molecular clouds can also be the site of
reprocessing, scattering and reflection of hard photons  from X--ray sources
located inside or outside the   clouds themselves.
The strong X--ray emission in the 6.4~keV line from
Sgr~B2 has been explained by
Koyama et al. (1996a) and by Sunyaev \& Churazov (1996)  with the reflection
of hard X--rays coming from the GC during a past outburst from Sgr~A*.
Also our data require the addition of a 6.4~keV line
to better fit the spectrum from Sgr~B2,
but we cannot claim the prevalence of the 6.4~keV fluorescent line
with respect to the 6.7 keV iron line as  in Murakami et al. (2000).
It is
interesting to note that X--ray emission from ONSs might have the right
properties to explain the observed fluorescent iron line.
If the source of
hard photons is embedded in the cloud, the luminosity required to produce
the flux in 6.4 keV line reported by Koyama et al. (1996), F$_{6.4}\sim
1.7\times 10^{-4}$~photons~cm$^{-2}$~s$^{-1}$,
is given by  $L_8\approx 10^{36} (0.1/\tau_{\rm T})$~erg~s$^{-1}$, where
$L_8$ is the source luminosity at 8 keV in the 8-keV-wide energy band
(Sunyaev \& Churazov 1998).
Accreting ONSs emit $\sim 1.5\times 10^{35}$~erg~s$^{-1}$, about 10\%
of the their total (unabsorbed) luminosity, in the same band. Since
the Thomson depth in Sgr~B2 is $\sim$0.7, this is in gross agreement with
the required value, taking into account that the cloud material absorbs
up to $\sim 50-60\%$ of the radiation at those energies. Emission
of continuum photons inside the cloud together with the derived value
of the scattering depth also provides an estimate of the 6.4 keV line
equivalent width, EW~$\sim \tau_{\rm T}\sim$0.7~keV (see again
Sunyaev \& Churazov 1998), which is close to the observed one.

\section{Conclusions}

The BeppoSAX survey of the Giant Molecular Clouds
Sgr~B, Sgr~C and Sgr~D in the GC region has allowed the detection of
diffuse X--ray emission from several radio sources,
and the discovery of an unresolved source, SAX~J1748.2--2808, probably
associated to a group of young stellar objects also observed as
an IRAS and maser source, or to a wind-shocked giant molecular core.  

Sgr~B2 is the strongest diffuse X--ray source. Our spectral results
are slightly different from those previously reported by
Koyama et al. (1996a). This can probably be ascribed to the
different spatial and spectral resolution of the instruments, as well as
to the   differences in the background subtraction.
We also provide an alternative explanation for the X--ray emission
from the Sgr~B2 molecular cloud, in terms
of accretion of the dense molecular cloud matter on
the surface of old isolated neutron stars.
This scenario can explain the observed luminosity from  Sgr~B2,
as well as the 6.4~keV iron line
emission, if the accretion takes
place on a small fraction of the neutron stars surface,
suggesting the presence of a non negligible  magnetic field
in old neutron stars.

We detected X--rays also  from the H\,{\sc ii} regions in
Sgr~D  and Sgr~C, and from part of the
non--thermal radio filament in the Sgr~C.
Unfortunately, a detailed spectral study of these features is
not possible with the limited statistics of the present data.
All we can say is that the emission spatially correlated with the
radio structures is more prominent in the soft band below 5 keV.
The X--rays from H\,{\sc ii} regions  could
be explained as  the integrated emission   from many protostars
embedded in these star-forming regions.

Finally, we detected  emission spatially correlated with 
the Southern rim of
the   SNR  G1.05--0.15 in Sgr~D.
If confirmed, this is the first detection in the X--ray range for this
SNR.

\begin{acknowledgements}
The \sax\ satellite is a joint Italian-Dutch programme.
L. Sidoli acknowledges an ESA Research Fellowship.
We thank Silvano Molendi and Giorgio Matt for
help with the data analysis, 
Fulvio Melia,  Michael Fromerth 
and Tim Oosterbroek for interesting discussions.
We are grateful to Joseph Lazio for providing the VLA radio data.
This research has made use of the ESO/ST-ECF Science Archive
(available at {\em http://archive.eso.org}), 
of the SIMBAD
database operated at Centre de Donn\'ees astronomiques in Strasbourg,
and of the High Energy Astrophysics 
Science Archive Research Center Online Service, 
provided by the NASA/Goddard Space Flight Center.
\end{acknowledgements}


\end{document}